\documentstyle[sprocl,epsfig,wrapfig]{article}

\title{Study of T-odd Quark Fragmentation Function
in $Z^0\to 2$-jet  Decay }
\author{A.V.Efremov \footnote{ Supported by RFBR under the
Grant 96-02-17631.}, O.G.Smirnova and \underline{L.G.Tkatchev}}
\address{\it JINR, Dubna, 141980 Russia }
\begin{document}
\maketitle

\abstracts{The first probe of the correlation of the T-odd one-particle
fragmentation function responsible for the left--right asymmetry
of fragmentation of a transversely polarized quark and an antiquark is done
by using the 1991-95 DELPHI data for $Z\to 2$ jet decay. Integrated over
the fraction of longitudinal and transversal momenta, this correlation
is of 2.5 ppm order, which means order of 7\% for the analyzing power.  This
makes us hope to use certain effects in polarized DIS  experiments for
transversity measurement.}

\section{Introduction}
The study of spin effects in high-energy interactions provide a sensitive
test for models of strong-interaction dynamics and have produced a number of
surprises. The transfer of nucleon polarization to quarks is investigated in 
deep-inelastic polarized lepton -- polarized nucleon scattering experiments
\cite{rep}. The corresponding nucleon spin structure functions for the
longitudinal spin distribution $g_1$ and  transversal spin distribution
$h_1$  for proton are well known.  The {\it opposite} process, the spin
transfer from partons to a final hadron, is also of fundamental interest.
Analogies of $f_1,\ g_1$ and $h_1$ are functions $D_1,\ G_1$ and $H_1$,
which describe the fragmentation of a non-polarized quark into a non-polarized
hadron  and a longitudinally or transversely polarized quark into a
longitudinally or transversely polarized hadron, respectively \footnote{ We
use the notation of the work \cite{muldt}.}.

These fragmentation functions are integrated over the transverse momentum
$\vec k_T$ of a quark with respect to a hadron. With $\vec k_T$ taken
into account, new possibilities arise. Using the Lorentz- and
P-invariance one can write in the leading twist approximation write 8
independent spin structures\cite{muldt,muldz}. Most spectacularly it is
seen in the helicity basis where one can build 8 twist-2 combinations,
linear in spin matrices of the quark and hadron $\vec\sigma,\ \vec S$ with
momenta $\vec k,\ \vec P$.  Especially interesting is a new structure that
describes a left--right asymmetry in the fragmentation of a transversely polarized
quark:  
$ 
H_1^\perp \vec\sigma(\vec P\times\vec k_T)/P\langle k_T\rangle\ ,
$
where the coefficient $H_1^\perp$ is a functions of the longitudinal
momentum fraction $z$, quark transversal momentum  $k_T^2$ and 
$\langle k_T\rangle$ is an average of the
transverse momentum. 

In the case of fragmentation to a
non-polarized or a zero spin hadron, not only $D_1$ but also the $H_1^\perp$
term will survive. Together with its analogies in distribution functions
$f_1$ and $h_1^\perp$, this opens a unique chance of doing spin physics
with non-polarized or zero spin hadrons! In particular, since the $H_1^\perp$
term is helicity-odd, it makes possibile to measure the proton
transversity distribution $h_1$ in semi-inclusive DIS from a transversely
polarized target by measuring the left-right asymmetry of forward produced
pions (see~\cite{mulddis,kotz} and references therein).

The problem is that, first, this function is completely unknown both
theoretically and experimentally and should be measured independently. 
Second, one should keep in mind that the function $H_1^\perp$ is the
so-called T-odd fragmentation function: under the naive time reversal $\vec
P,\ \vec k_T,\ \vec S$ and $\vec\sigma$ change sign, which demands a
purely imaginary (or zero) $H_1^\perp$ in the contradiction with hermiticity.
This, however, does not mean the breaking of T-invariance but rather the
presence of an interference term of different channels in forming the final
state with different phase shifts, like in the case of the single spin
asymmetry phenomena~\cite{gasior}. A simple model for this function could
be found in~\cite{colnuc}. It was also conjectured~\cite{jjt} that the
final state phase shift can average to zero for a single hadron
fragmentation upon summing over unobserved states $X$. Thus, the
situation here is far from being clear.

Meanwhile, the data collected by DELPHI (and other LEP experiments) give a
unique possibility to measure the function $H_1^\perp$.  The point is that
despite the fact that the transverse polarization of a quark ( an antiquark)
in Z$^0$ decay is very small ($O(m_q/M_Z)$), there is a non-trivial
correlation between transverse polarizations of a quark and an antiquark in the
Standard Model:  $ C^{q\bar q}_{TT}={(v_q^2-a_q^2)/(v_q^2+a_q^2)} $, which
reaches rather high values at $Z^0$ peak: $C_{TT}^{u,c}\approx -0.74$ and
$C_{TT}^{d,s,b}\approx -0.35$.  With the production cross section ratio
$\sigma_u/\sigma_d=0.78$ this gives the value
$\overline{C_{TT}}\approx -0.5$ for the average over flavors.

The spin correlation results in a peculiar azimuthal angle dependence of
produced hadrons (the so-called "one-particle Collins asymmetry"), if the
T-odd fragmentation function $H_1^\perp$ does 
exist~\cite{colnuc,colpsu,colar}.  The first probe of it was done three 
years ago~\cite{delnote95} by using a limited DELPHI statistics with the result $
\left|\overline{H_1^{\perp}/D_1}\right|\le 0.3, $, as averaged over quark
flavors.

A simpler method has been proposed recently by an Amsterdam group
\cite{muldz}. They predict a specific azimuthal asymmetry of a hadron in a
jet around the axis in direction of the second hadron in the opposite jet
\footnote{ We assume the factorized Gaussian form of $k_T$ dependence
for $H_1^{q\perp}$ and $D_1^q$ integrated over $|k_T|$.}:
\begin{eqnarray}
{d\sigma\over d\cos\theta_2 d\phi_1}\propto (1+\cos^2\theta_2)\cdot
\left(1+ {6\over\pi}\left[{H_1^{q\perp}\over D_1^q}\right]^2
C_{TT}^{q\bar q}{\sin^2\theta_2\over
  1+\cos^2\theta_2}\cos(2\phi_1)\right)
\label{mulders}
\end{eqnarray}
where $\theta_2$ is the polar angle of the electron and the second hadron
momenta $\vec P_2$, and $\phi_1$ is the azimuthal angle counted-off the
$(\vec P_2,\, \vec e^-)$-plane.  This looks simpler since there is no
need to determine the $q\bar q$ direction.

\section{Event Selection and Measurements}
This analysis covered  the DELPHI data collected from 1991 through 1995.  All
particles were generically assumed to be pions. Only charged particles were
analyzed.  About 3.5 millions of $Z^{\circ}$ hadronic decays were selected
by using the standard selection criteria~\cite{DELPHI-90}.

Jets were reconstructed by the JADE algorithm with varying the parameter
$Y_{cut} = 0.08,\ 0.05,\ 0.03$ or $0.01$.  Only 2-jet events were retained
for the analysis with additional thrust value selection requirement either
$T\ge 0$ or $T\ge 0.95$. To get rid of low efficiency of the
end-caps of the detector, events with the polar angle of the sphericity
axis $|\cos\theta_{\rm sp}|\ge 0.90$ were cut off and tracks with
$|\cos\theta_{\rm tr}|\ge 0.98$ were rejected, too.  In addition, the
acollinearity of the two jets $\Delta \theta _{jj}^{max}$ was required to
be $\leq5^{\circ}$.  A leading particle in each jet was selected both 
positive and negative.

To study the  detector response,  a sample of Monte-Carlo events, generated
with JETSET and passed through the same analysis chain as the data, was
used. With these events, the correction factor
\begin{equation}
f_{\rm corr} = {N_{\rm generated}(\theta_2,\phi_1)\over
N_{\rm simulated}(\theta_2,\phi_1)}
\label{fcorr}
\end{equation}
was built for each bin in the azimuthal angle of the first leading
particle $\phi_1$ and in the polar angle of the leading particle from the
opposite jet $\theta_2$ (see Expr. (\ref{mulders})).

The true distribution was defined as $N_{\rm true}= f_{\rm corr}N_{\rm
raw}$ and histograms in $\phi_1$  for each bin of $\theta_2$ were fitted by
the expression~\footnote{ The term with $\cos\phi_1$ is due to the twist-3
contribution of usual one-particle fragmentation, proportional to the
$k_T/E$.}
\begin{equation}
P_0(1+P_2\cos2\phi_1 + P_3\cos\phi_1).
\label{fit}
\end{equation}

\section{Results and Discussion}
For raw data $P_2^{\rm raw}$ is positive ($\approx 0.02$) for
$\theta_2$ close to $90^\circ$ but it becomes negative (up to $-0.09$) for
$\theta_2$ close to $0^\circ$ and $180^\circ$.  The same property but with
a larger value of $P_2^{\rm sim}$ ($\approx 0.03$ in the vicinity of
$90^\circ$) is shown by MC-simulated events too. This feature is clearly
interpreted as a consequence of low efficiency of the
DELPHI detector in the end-cups region and of the polar angle cut-offs.

Indeed, track 1 is more close to the cone of the "dead zone" when the angle
$\phi_1$ is close to $180^\circ$ (for $\theta_2<90^\circ$) or to $0^\circ$
(for $\theta_2>90^\circ$), which decreases the number of events
at the ends of $\phi_1$-histogram and produces a negative value of $P_2$. In
contrast to this, the low efficiency between TPC-segments of the
detector decreases in the number of events in the center of the
$\phi_1$-histogram (near $90^\circ$) and produces a positive values of $P_2$.

The positivity area increases for stronger jet selection criteria (smaller
$y_{\rm cut}$ and larger $T$-cut) with more narrow jets, but the value of
$P_2$ decreases.

The $P_2^{\rm gen}$ for pure JETSET shows a weaker dependence on $\theta_2$
and is much smaller in magnitude. In the region
$45^\circ<\theta_2<135^\circ$ this parameter is zero within the error bars.
Therefore this region was considered as the most reliable for the determination
of $P_2^{\rm true}$.

\begin{wrapfigure}{R}{6.3cm}
\mbox{\epsfig{figure=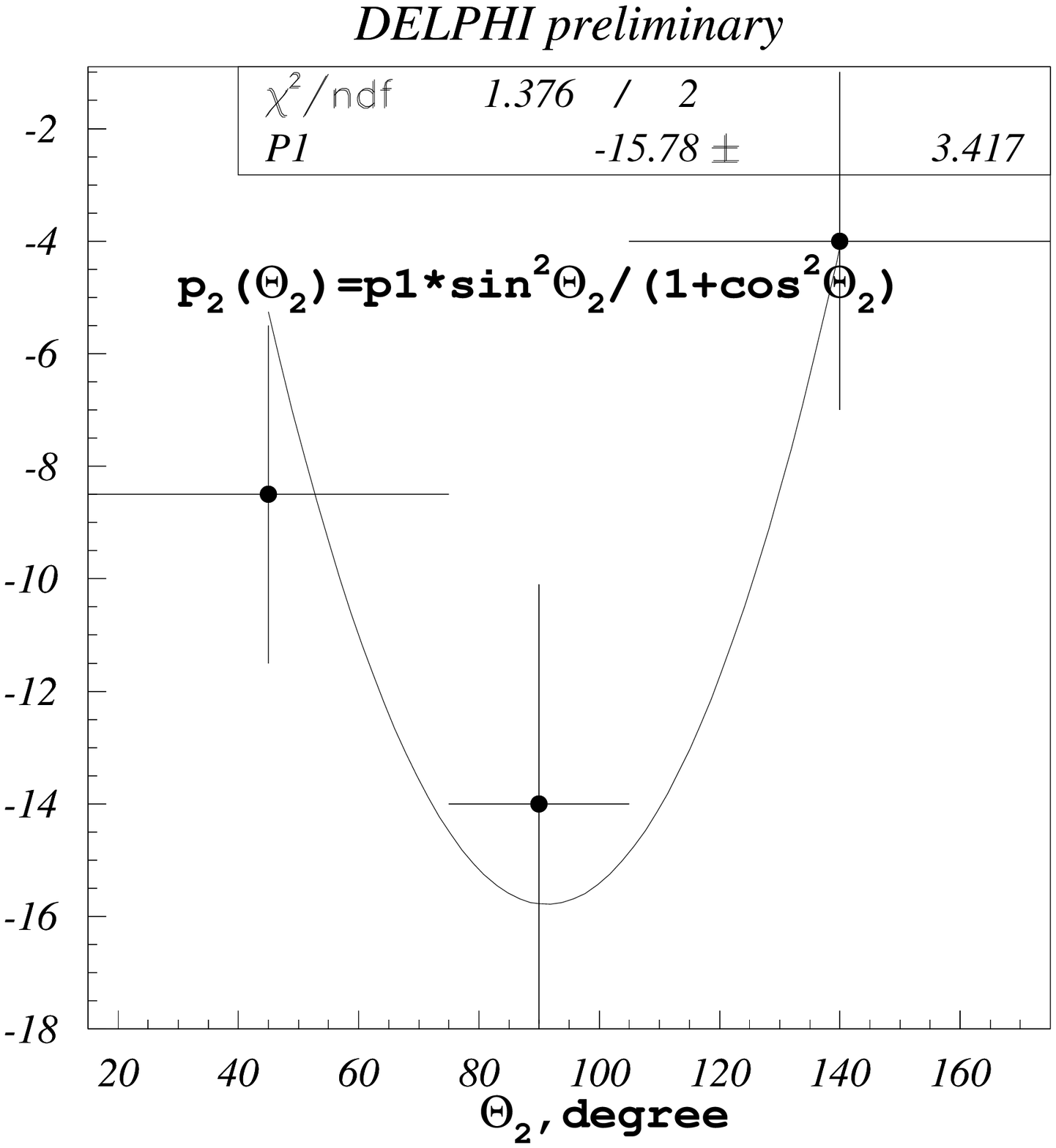,width=6.2cm,height=6.2cm}} \\
{\footnotesize Fig.1. The $\theta_2$-dependence of the
$P_2^{\rm true}$ (in ppm).}
\end{wrapfigure}
The best result for corrected data was obtained for  $y_{\rm cut}=0.03$ and
$T\ge 0.95$ selections. The value of $P_2^{\rm true}$ averaged over the
region $45^\circ<\theta_2<135^\circ$ and over quarks flavors with
$\overline{C_{TT}}\approx -0.5$ was found to be
\begin{equation}
P_2^{\rm true}=-0.0026\pm0.0018 \ .
\label{p2}
\end{equation}
The corresponding analyzing power according to Exp.(\ref{mulders}) is
\begin{equation}
\left|\overline{H_1^{\perp}\over D_1}\right| =6.3\pm1.7\% \ .
\label{apower}
\end{equation}

Regretfully, a rather small value of $P_2^{\rm true}$ and, especially, the 
fact that it was obtained effectively as a result of subtraction of much 
larger values of $P_2^{\rm raw}$ and $P_2^{\rm sim}$ do not allow us to 
consider the $\theta_2$-dependence of $P_2^{\rm true}$ seriously.  
Nevertheless, we risk to present this dependence in the whole interval of 
$\theta_2$ in Fig.1 with corresponding fit
$$
P^{\rm true}_2(\theta_2)=-(15.8\pm3.4){\sin^2\theta_2\over 1+\cos^2\theta_2}
\ ppm
$$ 
which increases the value of analyzing power (\ref{apower}) up to 
$12.9\pm1.4\%$. The distinction with (\ref{apower}) demonstrates, however, 
that systematic errors are by all means larger than the statistical ones 
and need further investigation.

To study this dependence in more detail, one has to increase the 
statistics.  It could be gained by inclusion not only the leading but also 
next-to-leading particles into study.  Also, the classical "Collins effect" 
should be investigated and confronted with the effect obtained.

In conclusion, we present some arguments in favor of a non-zero T-odd 
transversely polarized quark fragmentation function. The corresponding 
analyzing power could reach an order of 10 per cent, which makes us hope to 
use this effect for measiring of the transverse quark polarization in other 
hard processes. In particular, it can be done in the deep inelastic 
scattering for measurement of nucleon transversety distribution. Further 
increase of the accuracy and the investigation of systematic errors are 
required.

\medskip
We would like to thank G.Altarelli, D.Boer, A.Kotzinian, A.Olshevski and 
O.Teryaev for valuable discussions.  One of us (A.E) is obliged for support 
to TH CERN where a large part of this work was done.

\end{document}